\documentclass[12pt]{spieman}  
\usepackage{amsmath,amsfonts,amssymb}
\usepackage{graphicx}
\usepackage{setspace}
\usepackage{tocloft}
\usepackage{lineno}
\usepackage[labelfont=bf]{caption}
\usepackage{mathptmx} 
\usepackage[final]{changes} 
\definechangesauthor[name={Elena Carlson}, color=blue]{EC}

\DeclareUnicodeCharacter{22C6}{\ensuremath{\star}}
\DeclareUnicodeCharacter{03BE}{$\xi$}
\DeclareUnicodeCharacter{03B1}{$\alpha$}

\usepackage{graphicx} 
\usepackage{array}    
\usepackage{float}    
\usepackage{authblk}  
\usepackage{amsmath}  
\usepackage{hyperref} 
\usepackage{makecell} 

\usepackage{geometry} 
\usepackage[hang,flushmargin]{footmisc} 

\title{Push-broom Mapping of Galaxies and Supernova Remnants with the SPRITE CubeSat}

\author[a,b,*]{Elena Carlson}
\author[a,b]{Brian Fleming}
\author[a,b]{Yi Hang Valerie Wong}
\author[b]{Briana Indahl}
\author[b]{Dmitry Vorobiev}
\author[a,b]{Maitland Bowen}
\author[b]{Donal O'Sullivan}
\author[a,b]{Kevin France}
\author[c]{Anne Jaskot}
\author[d]{Jason Tumlinson}
\author[e]{Sanchayeeta Borthakur}
\author[f]{Michael Rutkowski}
\author[g]{Stephan McCandliss}
\author[d]{Ravi Sankrit}
\author[h]{John M. O’Meara}

\affil[a]{\small Department of Astrophysical and Planetary Sciences, University of Colorado, Boulder, Colorado, 80309}
\affil[b]{\small Laboratory for Atmospheric and Space Physics, 1234 Innovation Drive, Boulder, Colorado, 80309}
\affil[c]{\small Department of Astronomy, Williams College, Williamstown, MA 01267, USA}
\affil[d]{\small Space Telescope Science Institute, 3700 San Martin Drive, Baltimore, MD 21218, USA}
\affil[e]{\small School of Earth and Space Exploration, Arizona State University, 781 Terrace Mall, Tempe, AZ 85287, USA}
\affil[f]{\small Department of Physics and Astronomy, Minnesota State University, Mankato, MN 56001, USA}
\affil[g]{\small Center for Astrophysical Sciences, Department of Physics \& Astronomy, Johns Hopkins University, Baltimore, MD 21218, USA}
\affil[h]{\small W. M. Keck Observatory, 65-1120 Mamalahoa Hwy., Kamuela, HI 96743, USA}
\affil[*]{\small Corresponding author: Elenacarlson1@icloud.com}

\cftpagenumbersoff{figure}
\cftpagenumbersoff{table} 
\begin{document} 

\maketitle

\noindent
\begin{minipage}{\textwidth}
\footnotesize
\hrule height 0.2pt
\vspace{3pt}
\textbf{Citation:} Elena Carlson, Brian Fleming, Yi Hang Valerie Wong, Briana Indahl, Dmitry Vorobiev, Maitland Bowen, Donal O’Sullivan, Kevin France, Anne Jaskot, Jason Tumlinson, Sanchayeeta Borthakur, Michael Rutkowski, Stephan McCandliss, Ravi Sankrit, John M. O’Meara, ``Push-broom mapping of galaxies and supernova remnants with the SPRITE CubeSat,'' \textit{J. Astron. Telesc. Instrum. Syst.} \textbf{11}(4), 045001 (2025), doi: 10.1117/1.JATIS.11.4.045001.\\[4pt]
\textbf{Copyright:} © 2025 Society of Photo-Optical Instrumentation Engineers. One print or electronic copy may be made for personal use only. Systematic reproduction and distribution, duplication of any material in this paper for a fee or for commercial purposes, or modification of the content of the paper are prohibited.
\vspace{3pt}
\hrule height 0.2pt
\end{minipage}

\begin{abstract}

Supernovae (SNe) enrich and energize the surrounding interstellar medium (ISM) and are a key mechanism in the galaxy feedback cycle.  The heating of the ISM by supernova shocks, and its subsequent cooling is of critical importance to future star formation.  The cooling of the diffuse shock-heated ISM is dominated by ultraviolet (UV) emission lines.  These cooling regions and interfaces have complex spatial structure on sub-parsec scales.  Mapping this cooling process is an essential part of understanding the feedback cycle of galaxies, a major goal of the 2020 Astrophysics Decadal Survey.  The Supernova remnants and Proxies for ReIonization Testbed Experiment (SPRITE) 12U CubeSat Mission will house the first long-slit orbital spectrograph with sub-arcminute angular resolution covering far ultraviolet wavelengths (FUV; 1000 - 1750\textup{~\AA}) and access to the Lyman UV ($\lambda$ $<$ 1216\textup{~\AA}).  SPRITE is designed to provide new insights into the stellar feedback that drives galaxy evolution by mapping key FUV emission lines at the interaction lines between supernova remnants (SNRs) and the ambient interstellar medium (ISM).  SPRITE will also measure the ionizing escape from $\sim$ 50 low-redshift (0.16 $<$ z $<$ 0.4) star-forming galaxies.  Current models predict SPRITE capable of detecting strong O VI, O IV], and C IV emission lines with angular resolution from 10 - 20 arcseconds.  The SPRITE SNR survey will use push-broom mapping of its long-slit on extended sources to produce the first large sample of sub-arcminute 3D data cubes of extended sources in the FUV. In this paper, we present simulated SPRITE observations of Large Magellanic Cloud (LMC) SNRs to demonstrate the efficacy of the SPRITE instrument ahead of launch and instrument commissioning. These models serve as critical planning tools and incorporate the final pre-flight predicted performance of the instrument and the early extended source data reduction pipeline.
\end{abstract}

\keywords{astronomy, astrophysics, ultraviolet spectroscopy, supernova remnants, ionization, emission lines, interstellar medium}

\begin{spacing}{2}   

\section{Introduction}
\label{sect:intro}  

\subsection{Science Overview}

A majority of supernovae (SNe) form due to the core-collapse of high mass (M $>$ 8 $M_\odot$) O and B stars, and the remaining are due to the thermonuclear explosion of white dwarf stars.  Core-collapse SNe are closely connected to the stellar life-cycle as they explode in surroundings impacted by their progenitors.\cite{impactMassLoss}  During their lifetimes, O and B stars strongly influence the interstellar medium (ISM), injecting energy and momentum into the surrounding gas and dust via stellar winds, and emitting large quantities ($\sim$ $10^{53}$ $s^{-1}$) of ionizing radiation that ionize the ISM \cite{Leitherer_1995, Sternberg_2003, annurev:/content/journals/10.1146/annurev-astro-082812-140956}.

SNe are a major source of energy, heavy elements, and hot gas in the interstellar medium.  Forward and reverse shock waves heat the ISM, with the majority of the resulting ultraviolet (UV) emission features produced by the subsequent cooling of the heated gas \cite{2009IAUS..256..443W}.  Star-burst driven outflows result in gas being driven from the star-forming disk, heated from shocks and photoionization, and potentially escaping the galaxy disk \cite{Heckman_2017, annurev:/content/journals/10.1146/annurev-astro-091916-055240, galaxies6040138}.
SNe also provide a mechanism for clearing neutral hydrogen gas and dust, to create low density escape channels for Lyman continuum (LyC) photons \cite{leitet2013}. Unfortunately, most prior ultraviolet-sensitive instruments capable of detecting this emission have been designed as point-source spectrographs, such as the Cosmic Origins Spectrograph (COS) on the Hubble Space Telescope (HST). Those that have had imaging capability, such as the Spectroscopy of Plasma Evolution from Astrophysical Radiation (SPEAR) \cite{Edelstein_2006} or HST-STIS, have either lacked the resolution to track fine features (SPEAR), or a field-of-view too small for effective extended source mapping (STIS). The Supernova remnants and Proxies for ReIonization Testbed Experiment (SPRITE) \cite{Fleming22} is designed to bring effective imaging spectroscopy in the far-ultraviolet (FUV) to orbit for the first time. 

\begin{figure}[H]
\includegraphics[width=\textwidth]{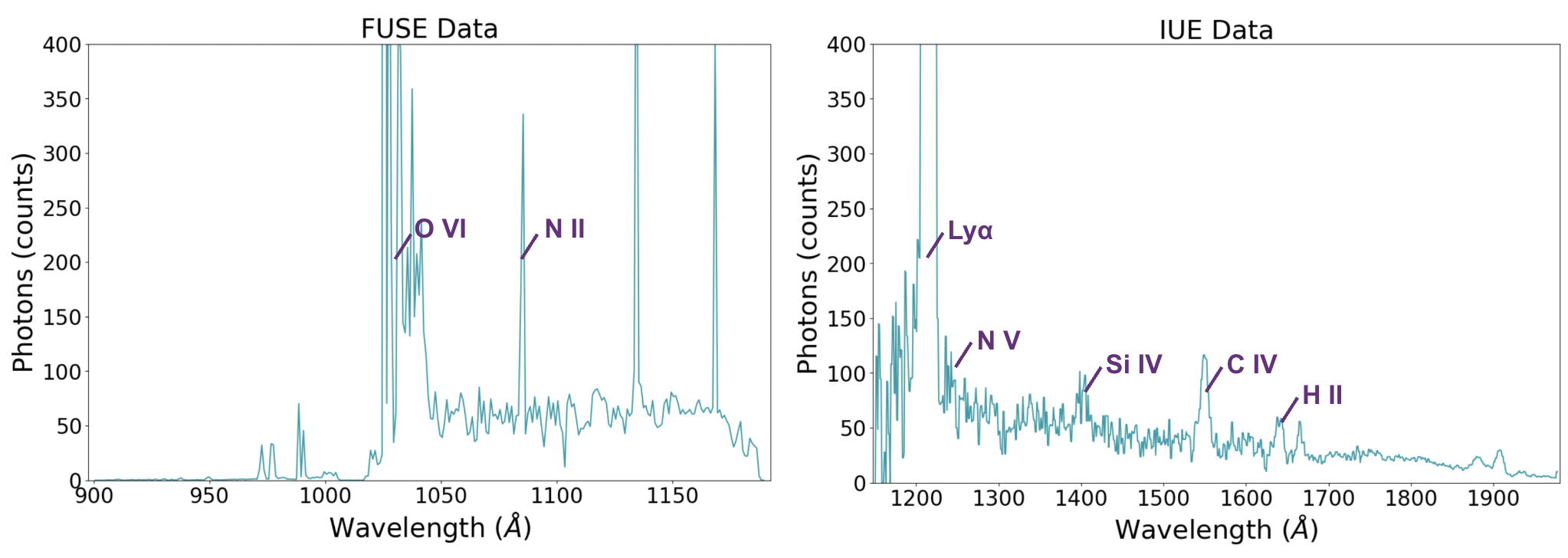}
\caption{FUSE and IUE spectra for N132D using separate pointings 1 and 2 respectively (see Section 4.4 for location in SNR). Prior to SPRITE, only HUT and SPEAR have been able to obtain spectra over the combined FUSE and IUE bandpass.}
\label{sep_pointing_specs}
\end{figure}

Some primary emission features observed in studies by the Far Ultraviolet Spectroscopic Explorer (FUSE) and International Ultraviolet Explorer (IUE) include O VI (1032, 1038\textup{~\AA}), N V (1238, 1242\textup{~\AA}), and C IV (1548, 1550\textup{~\AA})\cite{Sankrit_2007}.  Figure \ref{sep_pointing_specs} highlights FUV emission lines across the SPRITE waveband for two separate single-point regions within Large Magellanic Cloud (LMC) N132D, which has been chosen as a potential commissioning target for SPRITE due to the well-documented FUSE and IUE data on the supernova remnant (SNR).  When SNe-driven, these UV emission lines typically arise downstream from the SNe shock front and therefore produce a range of diagnostics about material being encountered and the physical processes involved \cite{Blair2017}.  O VI, in particular, is a unique tracer of warm shocked gas, as it arises shortly after the transition from non-radiative to radiative shocks at temperatures of $\sim$ 300,000 K, a level rarely reached by photoionization \cite{2003ApJ...586.1179D}.  This shock tracer is used in studies of winds from massive stars in the Large and Small Magellanic Clouds (LMC, SMC) to probe the high end of the gas temperature range throughout a wind \cite{Bianchi, Crowther_2000, Fullerton_2000, Massa_2000}.  Significant differences between wind properties for otherwise similar stars in the SMC and LMC have been confirmed using the O VI tracer, indicating differences in stellar evolution fueling the ISM \cite{Moos_2000}.

The extent to which gas and dust are produced post-SNe is a field of active study \cite{Slavin_2020, 2025arXiv250611931O, 2003Natur.424..285D}; 3D data cubes (X,Y,$\lambda$) produced from SPRITE data will constrain the timescales it takes retained gas to cool, and the potential for induced star formation.  Cooling rates can be derived by comparing to shock models and the spatially and spectrally resolved ion maps of these key UV diagnostics \cite{10.1111/j.1365-2966.2008.13121.x, 1980ARA&A..18..219M}.
 The relative spatial positions of ion emissivities are used to determine dust destruction and carbon depletion rates, as well as how shock velocities are moving through gas and dust near the progenitor.  These spatially resolved emission maps can also provide information about ion and electron temperature within the shock and surrounding dust clouds \cite{France_2011, 1983ApJ...275..636R}.
The SPRITE SNR survey will map supernova remnants and regions of star formation in the Milky Way, LMC, and SMC, with an additional survey of local galaxies to observe star-formation and nebular emission across galactic-structures.


The SPRITE maps will generate data on scales ranging from 2 - 4 pc (SMC/LMC) to 100 - 200 pc (e.g. M33) across the SPRITE bandpass (1000 - 1750 \AA ).  This technique will provide spectra of extended contiguous spatial regions on the sky and provide a fuller picture compared to past FUV missions such as FUSE \cite{10.1117/12.394015}, Hopkins Ultraviolet Telescope (HUT) \cite{1992ApJ...392..264D}, Hubble Space Telescope (HST), \cite{2012ApJ...744...60G} and IUE \cite{10625}, which obtained spectra only of smaller regions covered by single pointings.

The method of pushbroom mapping is a technique similar to that used by the SPEAR Mission \cite{Edelstein_2006}, though will be more focused and higher resolution.  SPEAR surveyed $\sim$ 80$\%$ of the diffuse FUV sky at a spectral resolving power of $\lambda$/$\Delta$$\lambda$ $\sim$ 550 at an imaging resolution of $\sim$ 5 arcminutes. SPRITE, by contrast, will observe approximately 50 discrete SNRs and 6 - 10 local galaxies at an angular resolution of 10 - 20 arcseconds and 3 \AA\ spectral resolution. The SPRITE raw data products and tools for data cube construction will be archived on the Mikulski Archive for Space Telescopes (MAST) beginning one year after the start of the nominal science mission. 


\subsection{The SPRITE Science Surveys}


SPRITE will collect spectral data between 1000 and 1750\textup{~\AA}, making it the first satellite to employ a push-broom technique in the FUV with access to the Lyman UV and sub-arcminute imaging resolution along a slit.  The science instrument will survey high energy emission from diffuse gas to study star formation feedback and its effects on the ISM.

This will be accomplished through three primary surveys.  [1] The SPRITE Ionizing Radiation Emitter Survey (SPIRES) will measure the ionizing escape from $\sim$ 50 low-redshift (0.16 $<$ z $<$ 0.4) star-forming galaxies. At z $\sim$ 0.3, these galaxies will be essentially point sources at the $\sim$ 10 arcsecond angular resolution of SPRITE, with the ionizing portion of the spectrum ($\lambda$ $<$ 912\textup{~\AA}) redshifted into the SPRITE bandpass. This survey is only briefly mentioned in this paper, as it does not require mapping techniques or extended source data processing. For the other two science surveys, [2] the SPRITE Galaxy Survey (SPRIGS) and the [3] SPRITE Topographic Emission Mapping Survey (STEMS), SPRITE will employ push-broom mapping to survey emission from star-forming regions and shocked gas around supernovae in the Milky Way, Magellanic Clouds, and in nearby (z $<$ 0.01) galaxies.  Push-broom mapping involves stepping SPRITE's long slit along a spatial region to create a 3D data cube of recorded flux values for specified right ascension, declination, and wavelength (X, Y, $\lambda$) in the field of interest.  Both of these surveys will produce an unprecedented ultraviolet picture of star-formation and feedback for comparison with existing data from ground-based, integral-field spectrographs and other surveys.  This paper focuses on the SPRIGS and STEMS surveys, with a companion paper by (Wong et al., in prep) focusing on the search for ionizing radiation with the SPIRES data.

\section{Instrument Overview}

SPRITE is a 12U (393 x 226 x 226 mm) NASA-funded CubeSat and the first NASA-funded orbital astrophysics spectrograph that is sensitive to the FUV since HST-COS \cite{2012ApJ...744...60G}.  Unlike past satellites only capable of studying a singular point source at a time, SPRITE's long-slit will collect more data in a shorter number of scans, 
providing a measure of the angular extent and profile of the FUV emission in a given extended source. The SPRITE program was designed both as a science mission as well as a testbed for enabling technologies for the Habitable Worlds Observatory (HWO), including advanced mirror coatings and photon-counting detector systems \cite{Bowen24}. 

\begin{figure}[H]
\includegraphics[width = \textwidth]{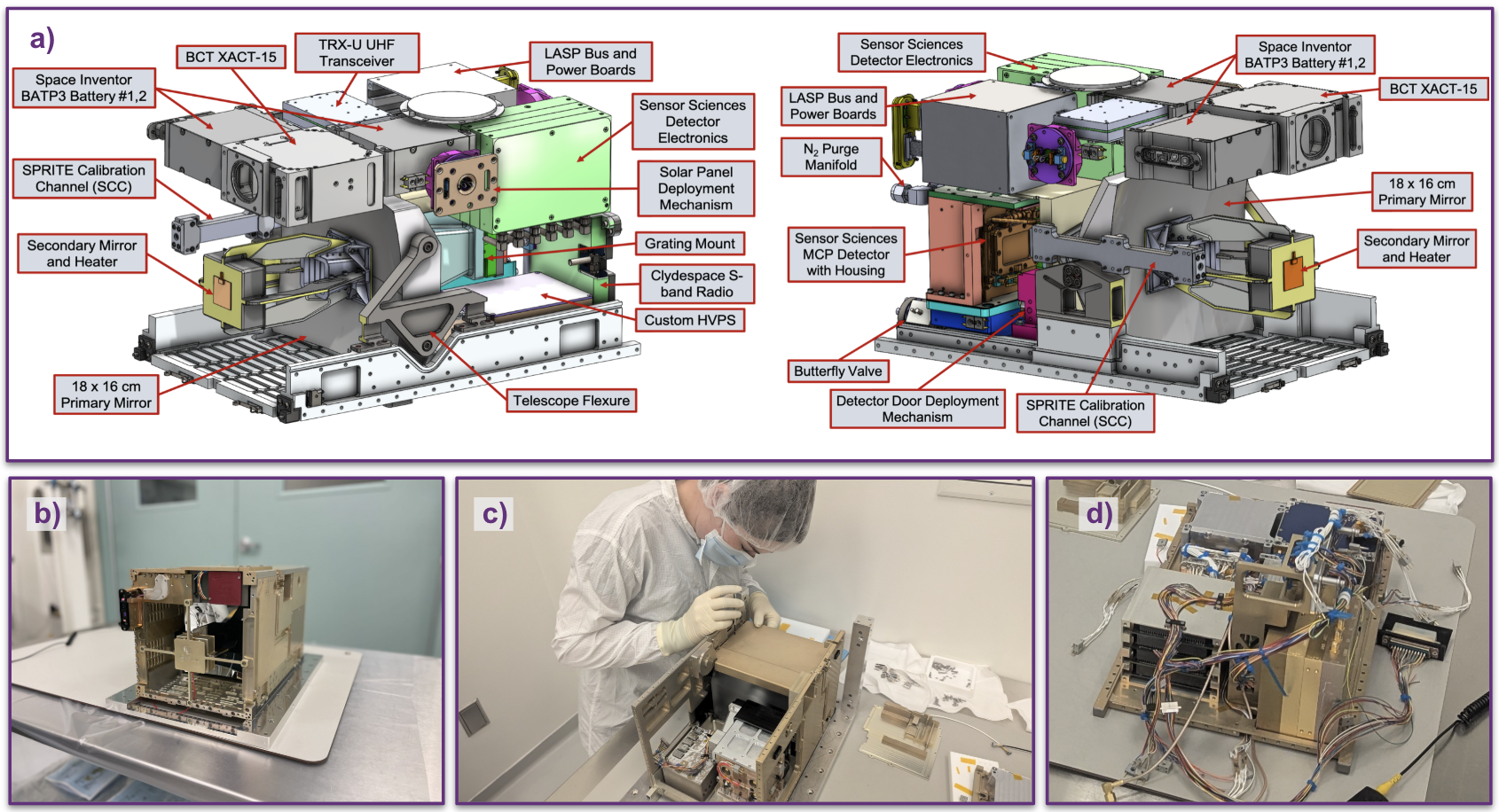}
\caption{Figure a) A rendering of SPRITE’s components inside of the chassis walls, including the spectrograph, telescope,
electronics, and spacecraft avionics; b) An image of the assembled satellite showcasing the telescope; c) SPRITE personnel assembling the base-plate with the science instrument; d) An image of SPRITE electronics.}
\label{cad_images}
\end{figure}

SPRITE consists of a 16 x 18 cm rectangular telescope coated at NASA Goddard Spaceflight Center (GSFC) with both conventional lithium fluoride protected aluminum (LiF+Al; primary mirror) and new enhanced reflectivity lithium fluoride protected aluminum (``eLiF''; secondary and fold mirror) coatings \cite{Luis22}. A LiF-based coating is essential for efficient throughput from 1000 -- 1150 \AA , however LiF is hygroscopic and slowly degrades with exposure to humidity. While prior programs such as FUSE successfully protected the optics with dry N$_{2}$, such protection is far harder for a CubeSat given the need for a rideshare launch. It is also more challenging for a large program such as HWO, as the quantity of N$_{2}$ required to protect a $\sim$ 6-meter primary mirror may cause human safety concerns. To mitigate this hygroscopicity, each of these SPRITE mirrors is protected from humidity exposure with an ultra-thin ($<$ 5 nm) layer of magnesium fluoride applied via atomic layer deposition at the NASA Jet Propulsion Laboratory \cite{2017ApOpt..56.9941F}. This has been shown to reduce hygroscopic degradation by at least an order of magnitude \cite{Luis22}, with these same coatings developed for SPRITE now employed on the Aspera SmallSat. These advanced coatings enable the SPRITE science program as planned, as several mirror reflections are required for sufficient aberration control in the limited SPRITE volume for large field-of-view long-slit spectroscopy over a large bandpass. Conventional LiF+Al on all optics, especially if degraded by humidity exposure, would have far lower throughput over multiple reflections than the more robust SPRITE mirrors. 

\begin{figure}[H]
\includegraphics[width = \textwidth]{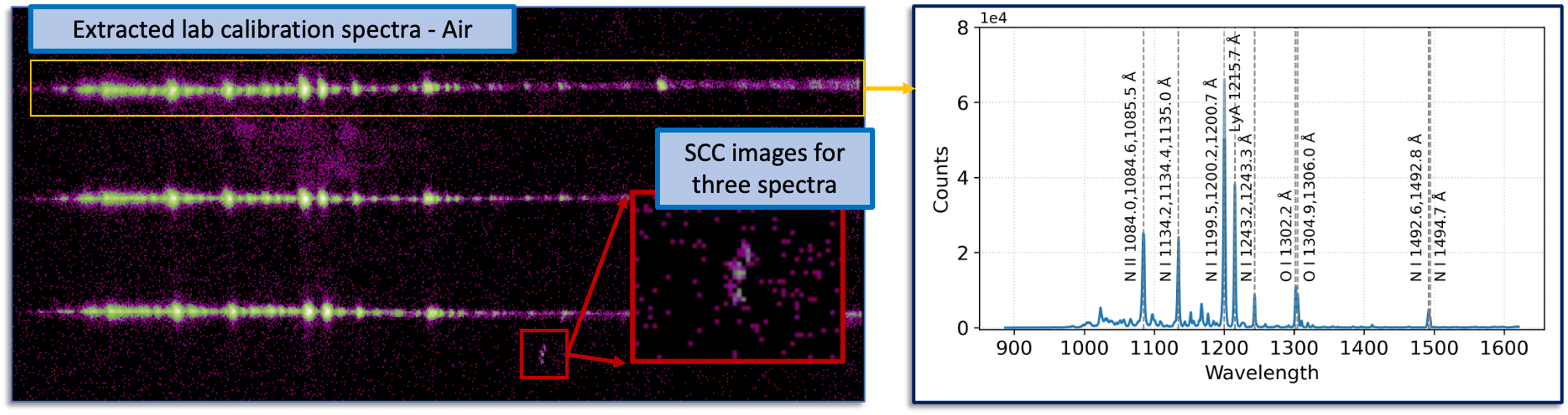}
\caption{Figure (Left) two-dimensional (raw) image from the SPRITE detector of a point-source lab calibration lamp spectra showing excited transitions of room air. The inset shows the three corresponding images through the SCC, tracking the distance between the slit regions. (Right) the extracted spectrum with lines labeled. These calibration images are used to establish the pre-flight imaging and spectral resolution, as well as scattered light background.}
\label{fig-SCCspectra}
\end{figure}

The telescope focuses light onto an 1800 $\times$ 10 arcsecond slit, with the diverging beam post-slit dispersed and re-focused by a custom holographically ruled blazed grating from Horiba Jobin-Yvon. The grating was originally coated in the same ALD-protected eLiF as the secondary and fold mirrors, however a late opportunity to change to a newer xenon-enhanced LiF+Al process (``XeLiF'') at GSFC was executed during SPRITE integration and testing, enabling the latest coating favored by HWO to be flight tested on-orbit for the first time \cite{10.1117/12.3021592}. The diffracted light is corrected for astigmatism by a cylindrical fold mirror and then imaged onto a 39 $\times$ 19 mm cross delay-line (XDL) microchannel plate detector (MCP) with a CsI photocathode.  Figure \ref{cad_images} shows a rendering of the SPRITE components inside the chassis walls, as well as images of the satellite during assembly.

To aid in tracking the performance of these experimental coatings in-flight, the spacecraft carries a complementary calibration channel designed to disentangle degradation of the optics from the detector system. This enables degradation of the optical coatings to be differentiated from contamination or detector-related degradation while on orbit. The SPRITE Calibration Channel (SCC) consists of a collimating tube with BaF$_2$ windows focusing a $\sim$ 2 degree FoV region onto an unused portion of the detector (Figure~\ref{fig-SCCspectra}).  BaF$_2$ and the MgF$_2$ mirrors of the SCC system are Technology Readiness Level (TRL) 9 with existing flight data under exposure to atomic oxygen, therefore cross-calibration measurements of standard stars between the SCC and primary science channel should provide a relative measure of the performance of the primary science channel mirror coatings to support HWO. The SCC has the added benefit of providing an ultra-low resolution imager for bright stars, aiding in star-tracker alignment during commissioning.

\renewcommand{\arraystretch}{1.5} 
\setlength{\tabcolsep}{8pt}       

\begin{table}[H]
  \centering
  \resizebox{\textwidth}{!}{
    \begin{tabular}{|c|c|c|}
      \hline
      \multicolumn{3}{|c|}{\textbf{SPRITE Instrument Parameters}} \\
      \hline
      \textbf{}  & \textbf{Primary Science Channel} & \textbf{SPRITE Calibration Channel} \\
      \hline
      \textbf{Purpose} & Imaging Spectroscopy & Calibration Imaging \\
      \hline
      \textbf{Aperture Size} & 
      \makecell{$18 \times 16$ cm \\ ($244.8\ \mathrm{cm}^2$ C.A.)} & 
      \makecell{$0.8 \times 0.8$ cm \\ ($0.64\ \mathrm{cm}^2$)} \\
      \hline
      \textbf{Angular Resolution} & $10''$ -- $22''$ & $2'$ -- $10'$ \\
      \hline
      \textbf{Field of View} & 
      \makecell{$1800'' \times 10''$ \\ $60'' \times 60''$ Center} & 
      $2^\circ \times 2^\circ$ \\
      \hline
      \textbf{Bandpass} & 1000 -- 1750~\AA & 1350 -- 2000~\AA \\
      \hline
      \textbf{Spectral Resolution} & 1.8~\AA & ------ \\
      \hline
    \end{tabular}
  }
  \setlength{\abovecaptionskip}{5pt}
  \caption{A summary of the instrument parameters for the SPRITE primary science and calibration channels. The calibration channel is designed to measure the 1400 - 2000\textup{~\AA} flux of a calibration target without using any of the LiF+Al-coated telescope optics, enabling separation of the detector and telescope degradation over time on orbit. The spectral resolution is given as an average over the bandpass.}
  \label{inst_params}
\end{table}

Within the 1800" of slit length there are three equally-spaced point source bulges; the top and bottom bulges are 30 x 30 arcseconds and the center bulge is 60 x 60 arcseconds.  These bulges provide added margin for pointing instabilities and reduced vignetting for point sources, while the remainder of the slit is optimized for extended object mapping. The pointing stability of the Blue Canyon Technologies XACT-15 module is estimated to be $\sim$ 8 arcseconds RMS, therefore for any given position on the sky, the subtended width will average the RSS of the slit width (10") and RSS pointing (8"), or $\sim$ 13 arcseconds. 

The SPRITE detector has been measured in the laboratory to have $\sim$ 0.1 event/cm$^2$/second, or $\sim$ 0.01 counts per 11 x 14 micron digital 
“pixel" per day.  This rate is expected to increase to $\sim$ 0.5 count/cm$^2$/s on-orbit (0.06 counts/pixel/day) due to the radiation environment.  Stray light from the optical system may increase this level further, as the compact nature of the SPRITE telescope and instrument is challenging to baffle \cite{2023SPIE12678E..0AB}, while the low scatter of geocoronal Lyman alpha off even a holographic grating is non-negligible for such a large field-of-view. In total, we expect between 2 - 10 events/cm$^{2}$/second in flight (0.25 - 1.3 events/pixel/day), depending on the position on the detector. This impacts the limiting sensitivity of SPRITE for the SPIRES survey, but has little impact on the expected signal-to-noise for the SPRIGS and STEMS observations, which are shorter exposure time observations.

\begin{figure}[H]
\includegraphics[width = \textwidth]{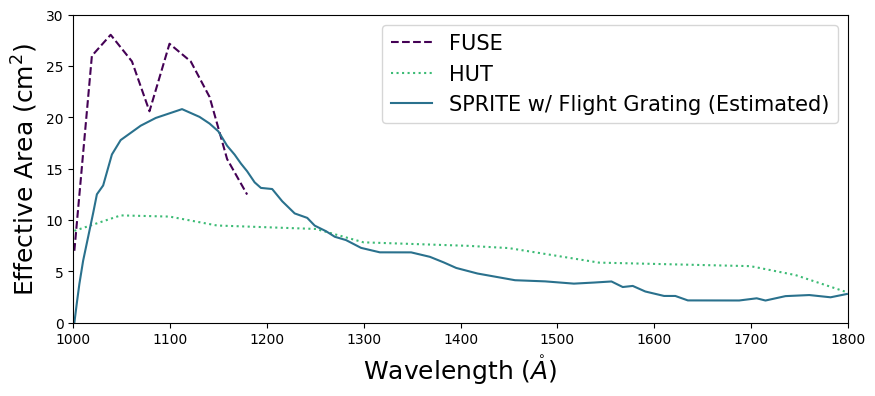}
\caption{SPRITE’s expected effective area in the case of using the flight grating is represented in solid blue.  The effective area of the two most relevant missions for comparison, HUT and a single LiF+Al channel of FUSE, are shown in the dotted green lines and dashed purple respectively.  The SPRITE effective area compares well with these previous larger and more expensive instruments.}
\label{eff_area}
\end{figure}

The peak effective area of SPRITE is projected to be 21 cm$^2$ using the best estimate measurements of each optical component pre-integration. The effective area is determined by the geometric collecting area, reflectance of each of its optics, absolute grating efficiency, and the quantum efficiency of the MCP detector. As the CubeSat must be delivered to the launch service provider 30 - 45 days before launch, and has no control over the launch environment in terms of cleanliness and the outgassing characteristics of the other payloads in the rideshare, it is expected that the on-orbit performance of SPRITE will be 30 to 50$\%$ lower than the pre-flight estimates. The science program is designed with margins compatible with these losses. For comparison, SPRITE’s peak effective area is 75$\%$ of FUSE's single channel effective area despite having twice as many optical reflections and less than 30$\%$ of the collecting area of a single FUSE channel. This highlights the advancements in mirror coating and detector technology enabled by NASA APRA, SAT, and Roman Technology Fellowship (RTF) Program investments over the last two decades. 


Laboratory testing of SPRITE has verified the angular and spectral resolution to be consistent with the values shown in Table \ref{inst_params} pre-launch.  The expected spectral resolution is between 1.5 and 3 \textup{\AA} with cross-dispersion resolution of 8 - 22 arcseconds.  This data will be published in the final preflight instrument paper, with a calibration spectrum of the emission features of air shown in Figure \ref{fig-SCCspectra}.  The final flight performance will be determined during commissioning.

\section{Modeling Effort}
\subsection{2D and 1D Modeling Process}

We model SPRITE 2D and 1D spectral data products using spectroscopic SNR observations from FUSE and IUE, specifically the Short-Wavelength Prime detector (SWP), downloaded from the MAST database.  Optical imaging data from the HST Advanced Camera for Surveys Wide Field Channel (ACS/WFC) were also used to produce extended source images and extract brightness profiles.  Models were produced using Python3 in conjunction with widely used packages such as Astropy\cite{astropy:2013, astropy:2018, astropy:2022}, Numpy \cite{harris2020array}, Scipy \cite{2020SciPy-NMeth}, and Matplotlib \cite{Hunter:2007}.

FUSE spectroscopic data were imported and binned to align with SPRITE’s spectral resolution of approximately 1 - 2\textup{~\AA}.  FUSE operates with a spectral resolution of $\sim$ 0.065\textup{~\AA} for a bandpass from 905 - 1187\textup{~\AA}, while IUE operates with a spectral resolution of 6 - 7\textup{~\AA} (in low resolution mode) for a bandpass from 1150 - 2000\textup{~\AA}. 

As neither instrument covers the full SPRITE bandpass, and rarely observed the same nebular regions in LMC/SMC SNRs, we approximate a potential SPRITE spectrum by combining different FUSE/IUE observations from the same SNRs, even if they were not co-spatial.  Overlapping wave values from the two different data sets were removed, favoring FUSE data due to the higher spectral resolution.  The data was then interpolated onto a common wavelength axis and converted to photon counts per\textup{~\AA} using the SPRITE effective area curve (see Figure \ref{eff_area}). The arrays were subsequently clipped at both ends to extend the exact width of SPRITE’s bandpass.

\begin{figure}[H]
\includegraphics[width=\textwidth]{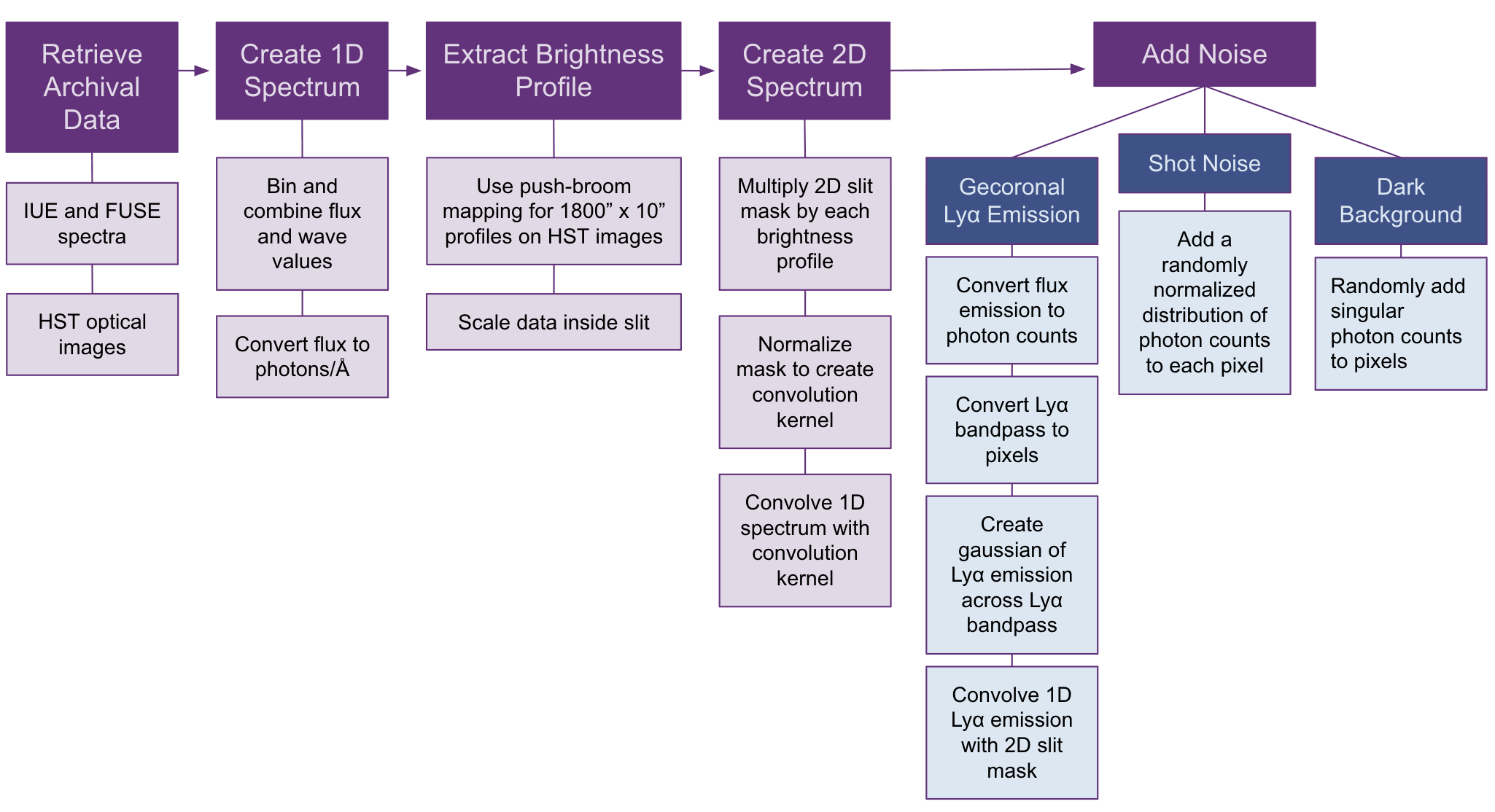}
\caption{The following block flowchart highlights the main steps of creating 1D and 2D spectral models.}
\label{flowchart}
\end{figure}

Following the creation of a 1D spectrum, the data were projected onto the x-axis of the detector via interpolation onto the SPRITE pixel plate scale.  We then approximate the extended emission by creating 1D profiles using optical and near-UV imaging from HST, and scale the combined FUSE/IUE spectrum by the brightness profile. This is unlikely to match the actual FUV brightness profile, however it does provide an approximation of the extent of the shocked gas.  Each brightness profile contained data along the 1800 arcsecond dimension of the slit spanning the length of the SNR.  The push-broom maps were then completed by extracting new brightness profiles for every 10 arcsecond step of the SPRITE slit.  The 1D spatial brightness profiles were converted to photon counts for a given exposure time and interpolated along the SPRITE y-axis pixel plate scale.  To scale the 1D brightness profile of the SNR within the slit, a target-size to slit-size ratio was applied to the data and a padding of zero-values was added to fill the rest of the detector y-pixels.

Once a 1D spatial profile was created, a 2D mask of the SPRITE slit was extracted from experimental detector data in the form of a .fits file.  This mask detailed the position and shape of the slit on the detector.  The mask was multiplied by the extracted brightness profile and normalized to create a convolution kernel.  The 1D spectral data was then convolved with the 2D convolution kernel to create a 2D spectrum.  This mathematical process is described below:

Creation of the convolution kernel ($K^\prime$) by multiplying the slit mask image (g) and brightness profile (b), then normalizing:
\begin{align}
K(x,y)  &= g(x,y) \cdot b(y) \\[1mm]
K'(x,y) &= \frac{K(x,y)}{\sum_{x}\sum_{y}K(x,y)}
\end{align}
\vspace{3mm}
Creation of the 2D spectrum by convolving the spectrum ‘f’ with a kernel ‘K’:
\begin{align}
I(x,y)= (K'\ast f)(x,y) \\[1mm]
f \longrightarrow f(x) \\[1mm]
K' \longrightarrow K'(x,y)
\end{align}

Noise from dark current background, shot noise, and geocoronal Lyman-alpha (Ly$\alpha$) emission was incorporated into the 2D profile.  Dark current was simulated by adding singular integer photon counts to randomly chosen pixels.  The number of photon counts added across the entire detector is determined by the product of the dark rate and exposure time.  Shot noise was included by adding a randomly normalized distribution of photon counts to each pixel. The contribution of instrument noise is considered negligible in this analysis, as the signal-to-noise ratio (S/N) is dominated by photon statistics.

Geocoronal Ly$\alpha$ noise was produced by starting with an expected value of geocoronal Ly$\alpha$ brightness \cite{2012AAS...21924118A}, and converting this value to photon counts.  All SPRITE observations will be carried out at night with an expected geocoronal Ly$\alpha$ background of 5000 Rayleigh in low earth orbit.  Other in-band geocoronal lines are negligible at night.  

Using the SPRITE x-axis pixel plate scale, the Ly$\alpha$\textup{~\AA} range was converted to pixels and used to create a Gaussian centered around a mean value equivalent to the geocoronal Ly$\alpha$ emission brightness.  An array of zero values was added on either side of the Gaussian to fill the remaining wavelengths of the spectrum.  Using the same process described mathematically above, a new convolution kernel was created from the 2D slit mask. The 1D geocoronal Ly$\alpha$ noise spectrum was convolved with this convolution kernel to create a 2D geocoronal Ly$\alpha$ noise array.  The 2D dark current background and shot noise arrays were added to this array to create a 2D noise array.  Finally, the 2D noise array and 2D SNR spectrum were added together and plotted to create a 2D simulated spectrum.  A summary of the modeling effort is outlined in the block flowchart from Figure \ref{flowchart}.

\begin{figure}[H]
\includegraphics[width=\textwidth]{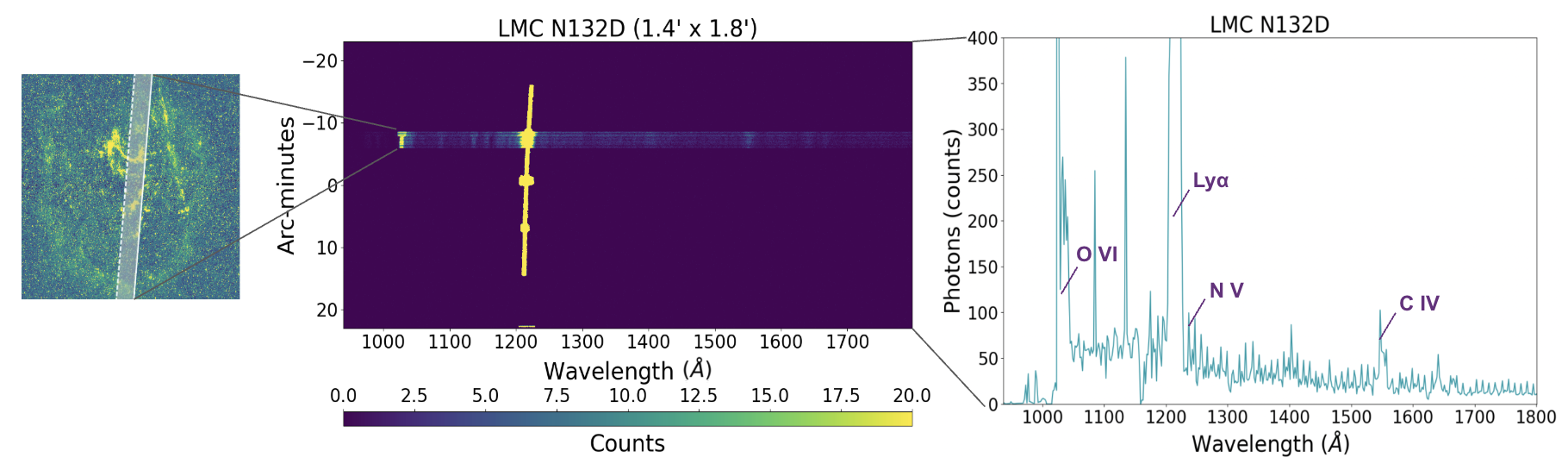}
\caption{The far left depicts supernova remnant N132D and an overlaid slit representing the 7th step in push-broom mapping the extended source.  The central figure depicts the corresponding simulated 2D spectrum for an exposure time of 3 kiloseconds.  The supernova remnant size of $\sim$ 1.4 x 1.8 arcminutes \cite{refId0} is depicted in the title.  The far right depicts the collapsed 1D spectrum with key emission features labeled. Imaging data was extracted from HST while spectral data was extracted from FUSE and IUE.}
\label{2d_spec}
\end{figure}

\subsection{Push-Broom Mapping}
\sloppy
Push-broom mapping describes the process of scanning an extended source using multiple steps of SPRITE’s long-slit in the dispersion direction.  Combining each of these spectral slices will allow for the production of a data cube that maps emission from a complete SNR source.  The SPRITE mission implements push-broom mapping because most prior UV spectrographs have been point source only, with SPRITE being the first orbital long-slit with sub-arcminute resolution.  Figure \ref{2d_spec} illustrates the result of modeling spectral and spatial data from a single push-broom scan.  Comparing modeled data from bright and well-known SNRs in the LMC and SMC, the number of push-broom steps is predicted to range anywhere from 1 to 60 for a given extended source, with most sources likely mapped in under 15 steps.  In the following section, Table \ref{snr_list} describes a list of 9 well documented SNRs and the predicted number of push-broom steps to scan the entire source.

Push-broom mapping allows for the production of a 3D data cube consisting of spatial (X,Y) as well as spectral ($\lambda$) information--essentially creating an image of the target at all resolved wavelengths.  SPRITE will observe up to 50 SNRs and create data cubes for each, with the data products archived on MAST alongside the relevant calibration data.  The exact pointing information for the spacecraft will result in a significant uncertainty as to the position of the slit, with the anticipated jitter of the Blue Canyon Technologies XACT-15 Attitude Determination and Control System (ADCS) on the order of the SPRITE angular resolution.  This will result in a blurring of the spectral maps, reducing the final angular resolution to $\sim$ 10 arcseconds as set by the slit width.  The degree of blurring will be determined by the Blue Canyon attitude control system in SPRITE.  Nevertheless, the profiles of the UV emission will still be the highest resolution accessible maps of these targets to-date at these wavelengths. The slit angle will be positioned to optimize the observing time of SPRITE, using as few push-broom steps as possible for a given source.
 
\begin{figure}[H]
\includegraphics[width=\textwidth]{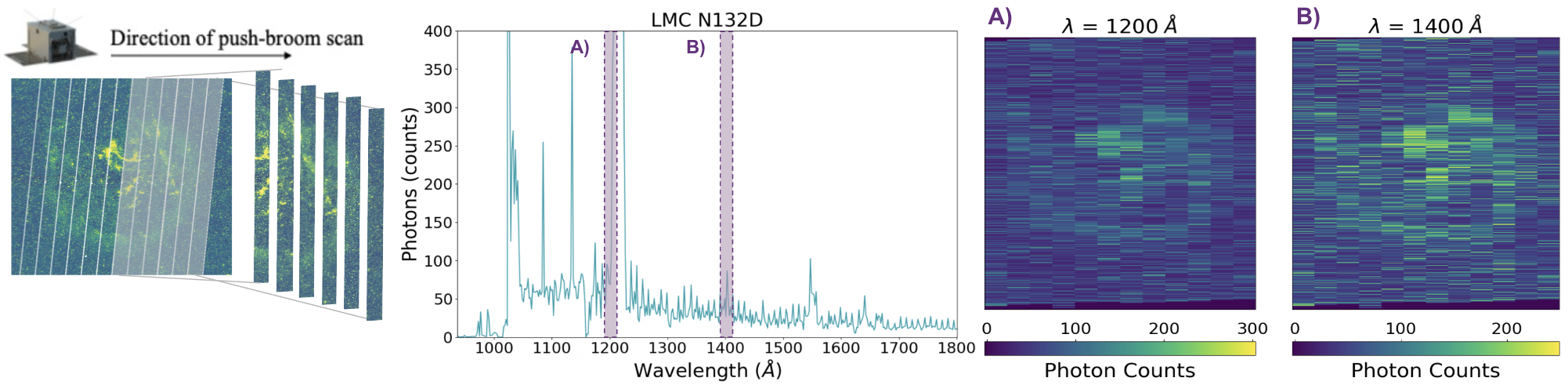}
\caption{The far left depicts supernova remnant N132D and an example of how SPRITE can use push-broom mapping to extract data from the SNR.  The central figure depicts the 1D spectral profile of a single slit for a 3 kilosecond exposure time, with two different chosen wavelengths highlighted in purple.  The right two figures depict a reconstruction of the SNR using brightness profiles extracted from the slit at 1200 and 1400\textup{~\AA} respectively.  Imaging data was extracted from HST while spectral data was extracted from FUSE and IUE.}
\label{reconstructed_image}
\end{figure}

A reconstructed SNR image at different wavelengths was modeled by interpolating each push-broom brightness profile onto the y-axis pixel plate scale. Each interpolated profile was extended 40 pixels in the x-direction to fill the width of the slit. The 2D arrays corresponding to each profile were conjoined and plotted to produce the SNR image reconstructed from the single slit models (see Figure \ref{reconstructed_image}).

\section{Commissioning Plan}
\begin{figure}[H]
\includegraphics[width=\textwidth]{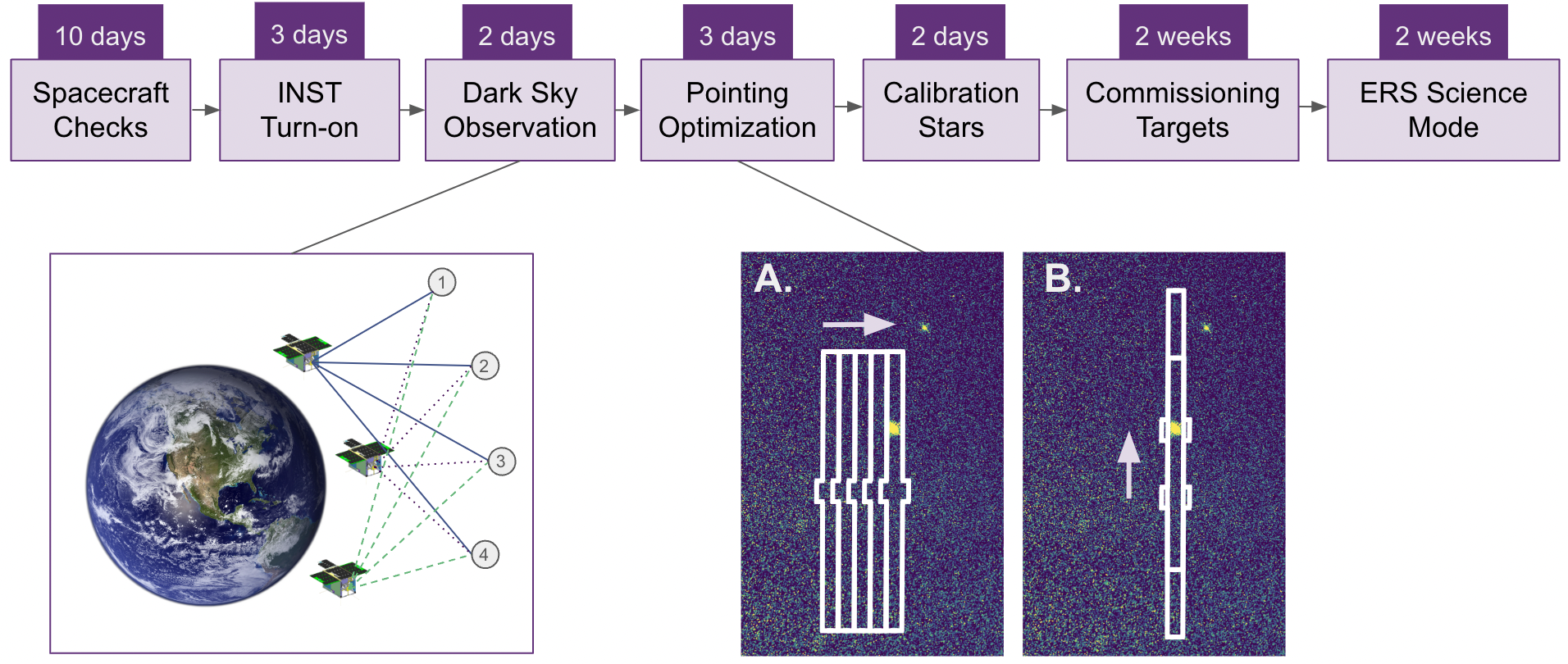}
\caption{The top chain diagram outlines the SPRITE commissioning plan with the optimal expected durations.  Unanticipated challenges from operations, pointing, and spacecraft control will almost certainly extend these times, as CubeSats cannot be operated with flight program efficiency.  The bottom left figure illustrates dark sky observation and how SPRITE will examine multiple different galactic latitudes from different orbital positions.  The bottom right illustrates pointing optimization and how SPRITE will utilize push-broom mapping capabilities to optimize calibration star pointing.  Part A shows how the slit will be moved over in ten arcsecond intervals until the star is found. Part B then illustrates how the slit will move perpendicularly to center the star within the slit.  The bottom right figure was created using HST optical imaging data of calibration star Hz4.}
\label{comm_plan}
\end{figure}

The previous section establishes a target that will be observed during SPRITE's commissioning phase.  After SPRITE's launch in 2025, approximately 2 weeks will be allotted towards spacecraft and instrument check-outs.  Following these crucial steps, two more days will be reserved for dark sky observation.  Here, SPRITE will examine regions free of bright sources at different galactic latitudes throughout an orbit.  This will allow the CubeSat to measure differences in the dark sky background due to changing orbital positions, and track potential scatter from the geocorona or relative Earth/limb angles.  Three days will then be spent optimizing pointing positions by utilizing SPRITE's push-broom capabilities to center calibration stars within the slit.  The spacecraft will move each pointing in a 10 arcsecond step raster scan, drawing out an increasingly large spiral pattern until the calibration star is found.  The slit will then move perpendicularly until the star is centered within the 60 x 60 arcsecond aperture.  Once a a star is centered, digital shims will be applied to the pointing routine for observing other targets. SPRITE will then observe and record spectral data for each calibration star.  The spacecraft will follow the calibration observations by observing early release/calibration targets consisting of known objects and pre-existing data.  This will include at least two LyC emitting galaxies selected with pre-existing HST-COS observations as part of the Low Redshift Lyman Continuum Survey \cite{Flury_2022},(Wong et al., in prep) as well as LMC/SMC SNRs with pre-existing FUSE and IUE observations of nebular emission.  The observation of targets with pre-existing data is an essential step in validating the SPRITE calibration and data processing routines.  Once the commissioning target observations are complete, SPRITE will transition towards nominal science operations for early release science (ERS) targets.  An overview of the commissioning plan is outlined in Figure \ref{comm_plan}.

\subsection{Calibration Target Selection}

From an established list of standard stars, data from HST, FUSE, and IUE provided by the MAST database were plotted as photon counts per \textup{\AA}, flux per \textup{\AA}, and S/N per \textup{\AA}.  The data were analyzed for two exposure times, 100 and 1000 seconds, to provide a lower and upper limit to the analysis.  Stars with plots showing a S/N per \textup{\AA} of 10 in at least one spectral line across the SPRITE waveband (1000 - 1750 \textup{\AA}) were chosen as calibration stars.  

Multiple limiting brightness factors were considered when choosing calibration targets.  Targets producing more than 100,000 counts per second were discarded for exceeding the software count rate limit.  If a star was too dim and did not reach the S/N minimum of 10 within the 1000 second exposure time limit, the star was also discarded from the calibration list.

As part of each calibration run, the star will nominally be placed at the center of the central 60 x 60 arcsecond bulge and observed from three locations at the exposure time listed in Table \ref{cal_targs}. To track the stability of the digital shimming of the spacecraft ADCS, the spacecraft will then step in each direction by 20 arcseconds, ultimately creating a 3 x 3 grid of observations. The count rates from each will be measured on the ground to verify that the nominal digital shims applied to the ADCS pointing still appear to fully place the star in the central bulge. 

For extended sources, two calibration stars will be used to calibrate the SPRITE roll angle.  For different extended sources, a different roll angle is required to maximize the amount of data recorded for a minimized number of push-broom steps.  A recalibration of the roll knowledge will also happen periodically to ensure the proper construction of the push-broom maps.  The satellite will then rotate until both stars are within the slit.

\begin{table}[H]
  \centering
  \resizebox{\textwidth}{!}{
    \begin{tabular}{|c|c|c|c|c|c|}
      \hline
      \multicolumn{6}{|c|}{\textbf{}} \\ 
      \multicolumn{6}{|c|}{\textbf{SPRITE Stellar Calibration Targets}} \\
      \multicolumn{6}{|c|}{\textbf{}} \\ 
      \hline
      \textbf{Target Name}  & \textbf{Right Ascension} & \textbf{Declination} &
      \textbf{Exposure Time} &
      \textbf{Signal-to-Noise} &
      \textbf{SCC Signal-to-Noise} \\
       & (h:m:s) & (deg:arcmin:arcsec) & (seconds) & (@ $\lambda$ = 1300\textup{\AA}) & (integrated across all $\lambda$) \\
      \hline
      Feige 34         & 10:39:36.71 & +43:06:10.1  & 100  & 47.3  & 13.6 \\
      Feige 110        & 23:19:58.39 & -05:09:56.1  & 100  & 30.4  & 8.6 \\
      G191-B2B         & 05:05:30.62 & +52:49:51.9  & 100  & 39.2  & 10.4 \\
      BD+33 2642       & 15:51:59.86 & +32:56:54.8  & 100  & 19.6  & 6.7  \\
      NGC 7293 Star    & 22:29:38.46 & -20:50:13.3  & 100  & 20.0  & 5.1  \\
      Hz 43            & 13:16:21.85 & +29:05:55.4  & 100  & 23.8  & 5.6  \\
      Hz 44            & 13:23:35.37 & +36:08:00.0  & 100  & 28.1  & 8.3 \\
      GD 246           & 23:12:21.6  & +10:47:04.3  & 100  & 20.1  & 5.4  \\
      AGK +81 266      & 09:21:19.18 & +81:43:27.6  & 100 & 28.8  & 9.5 \\
      GD 108           & 10:00:47.33 & -07:33:31.2  & 1000 & 31.5  & 7.5  \\   
      GD 50            & 03:48:50.06 & -00:58:30.4  & 1000 & 40.4  & 9.2  \\
      GD 71            & 05:52:27.62 & +15:53:13.2  & 1000 & 56.6  & 14.7 \\
      GD 153           & 12:57:02.32 & +22:01:52.6  & 1000 & 57.8  & 13.4 \\
      GD 659           & 00:53:17.43 & -32:59:56.5  & 1000 & 50.0  & 12.7 \\
      Hz 2             & 04:12:43.51 & +11:51:50.4  & 1000 & 22.6  & 5.6  \\
      Hz 21            & 12:13:56.42 & +32:56:30.8  & 1000 & 30.2  & 6.6  \\
      GRW+70 5824      & 13:38:51.77 & +70:17:08.5  & 1000 & 38.7  & 10.3 \\
      G93-48           & 21:52:25.33 & +02:23:24.3  & 1000 & 30.4  & 8.6 \\
      BPM 16274        & 00:50:03.65 & -52:08:15.5  & 1000 & 17.7  & 3.7   \\
      \hline
    \end{tabular}}
  \setlength{\abovecaptionskip}{5pt} 
  \caption{The table showcases a list of 19 selected calibration stars.  SPRITE only observes at orbital night, where the O I emission line is expected to be less than 100 Raleigh \cite{Hirschauer2024}. Column 6 specifies the Satellite Calibration Center (SCC) S/N integrated across the entire SPRITE bandpass.}
  \label{cal_targs}

\end{table}

\subsection{Commissioning Target Selection}
SNR commissioning targets were chosen on the basis of photon counts and S/N for an exposure time of 1000 seconds.  The SNRs with the highest S/N at key emission lines, such as O VI (1032, 1038\textup{\AA}), O IV] (1397, 1401), C IV (1548, 1550\textup{\AA}) were chosen as commissioning targets.  FUSE and IUE data gathered from the MAST database were used to analyze a list of LMC targets. The final selection included five SNRs, featured in Table \ref{com_targs}.

\begin{table}[H]
  \centering
  \resizebox{\textwidth}{!}{
    \begin{tabular}{|c|c|c|c|c|}
      \hline
      \multicolumn{5}{|c|}{\textbf{}} \\ 
      \multicolumn{5}{|c|}{\textbf{SPRITE Commissioning Targets}} \\
      \multicolumn{5}{|c|}{\textbf{}} \\ 
      \hline
      \textbf{Target Name} & 
      \textbf{Right Ascension} &
      \textbf{Declination} &
      \textbf{O VI Signal-to-Noise} &
      \textbf{C IV Signal-to-Noise} \\
      & (h:m:s) & (deg:arcmin:arcsec) & ($\lambda$ = 1032, 1038) & ($\lambda$ = 1545 - 1555) \\
      \hline
      N157B  & 05:37:46 & -69:10:28 & 16.21 & 0.54 \\
      N157C  & 05:35:60 & -69:12:00 & 12.72 & 0.66 \\
      N103B  & 05:08:59 & -68:43:35 & 9.65  & 0.89 \\
      N4D    & 04:53:10 & -66:54:48 & 12.79 & ---  \\
      N132D  & 05:25:04 & -69:38:24 & 10.09 & 1.6  \\
      \hline
    \end{tabular}
  }
  \setlength{\abovecaptionskip}{5pt} 
  \caption{The table showcases a list of 5 SNR commissioning targets. The S/N was calculated using an exposure time of 1000 seconds. C IV S/N was averaged over the specified wavelengths.}
  \label{com_targs}

\end{table}

\begin{figure}[H]
\includegraphics[width=\textwidth]{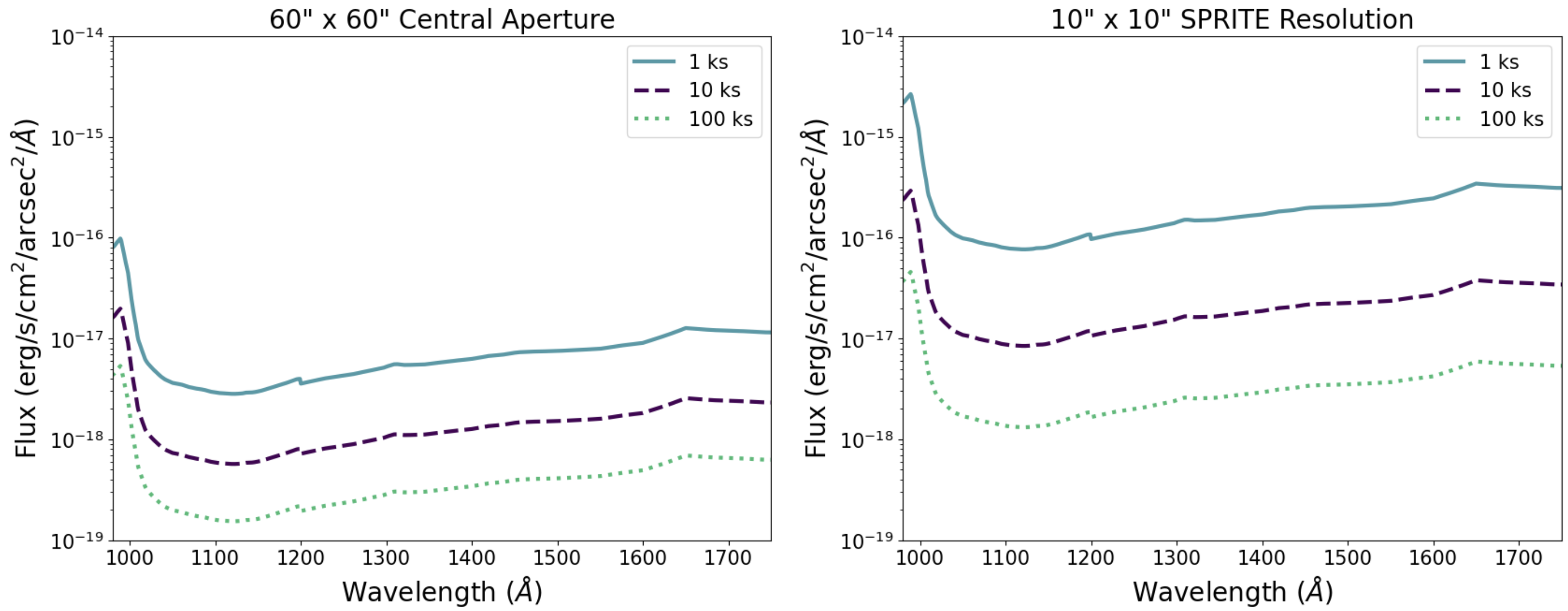}
\caption{The left-hand figure shows the minimum flux requirements for three separate exposure times to achieve a S/N of three across the central (60 x 60 arcsecond) aperture.  The right-hand figure shows corresponding minimum flux requirements at the maximum SPRITE resolution (10 x 10 arcsecond).}
\label{ext_source_sens}
\end{figure}

Figure \ref{ext_source_sens} illustrates SPRITE's sensitivity to various surface brightnesses as a function of 1000, 10,000, and 100,000 s exposure times.  SPRITE has roughly 2000 seconds of exposure time per orbit.
The plot emphasizes the flux required at each wavelength to achieve S/N = 3.  Flux values were calculated using the SPRITE detector background rate of 0.5 counts/cm$^{2}$/s \cite{10.1117/12.2561753}.  For SPRITE's large field of view, scattered geocoronal Ly$\alpha$ is expected to dominate at a rate of 3 - 5 counts/cm$^{2}$/s.  For this study, this has no bearing on our SNR results as it is photon statistics dominant. 

\subsection{Extended Source Pointing}
From a list of 59 SNRs in the LMC and 19 SNRs in the SMC \cite{refId1, refId2}, a modified list of nine science targets were chosen based on bright UV and optical imaging data.  These targets span a wide range of ages and progenitors, however the nature of the SNR in the Magellanic Clouds is not a factor in target selection because SPRITE is acting as a survey instrument. 
 Using this imaging data, the start and end slit position as well as roll angle were determined for different SNRs.

\begin{table}[H]
  \centering
  \resizebox{\textwidth}{!}{
    \begin{tabular}{|c|c|c|c|c|c|c|}
      \hline
      \multicolumn{7}{|c|}{\textbf{}} \\ 
      \multicolumn{7}{|c|}{\textbf{Large Magellanic Cloud}} \\
      \multicolumn{7}{|c|}{\textbf{}} \\ 
      \hline
      \textbf{SNR Name} & \textbf{Initial RA} & \textbf{Initial Dec} & \textbf{Final RA} & \textbf{Final Dec} & \textbf{Angle} & \textbf{Number of Pointings}  \\
      & (hr, min, sec) & (deg, arcmin, arcsec) & (hr, min, sec) & (deg, arcmin, arcsec) & (deg) &   \\
      \hline
      N157B     & 5:36:37 & -69:06:10 & 5:37:35 & -69:13:55 & 146.5 & 57 \\
      N63A      & 5:34:34 & -65:58:04 & 5:34:24 & -66:00:53 & 199.8 & 19 \\
      N132D     & 5:24:35 & -69:47:29 & 5:24:16 & -69:46:39 & 296.9 & 12 \\
      N49       & 5:27:10 & -66:09:45 & 5:27:02 & -66:10:47 & 218.0 & 9  \\
      N103B     & 5:09:30 & -68:30:31 & 5:09:34 & -68:30:34 & 97.8  & 3  \\
      DEM L71   & 5:06:52 & -67:58:16 & 5:06:44 & -67:58:54 & 229.8 & 7  \\
      B0519-690 & 5:17:26 & -68:54:21 & 5:17:29 & -68:53:56 & 33.0  & 4  \\
      B0540-693 & 6:39:55 & -69:27:32 & -------- & ---------- & 280.7 & 1 \\
      \hline
      \multicolumn{7}{|c|}{\textbf{}} \\ 
      \multicolumn{7}{|c|}{\textbf{Small Magellanic Cloud}} \\
      \multicolumn{7}{|c|}{\textbf{}} \\ 
      \hline
      \textbf{SNR Name} & \textbf{Initial RA} & \textbf{Initial Dec} & \textbf{Final RA} & \textbf{Final Dec} & \textbf{Angle} & \textbf{Number of Pointings}  \\
      & (hr, min, sec) & (deg, arcmin, arcsec) & (hr, min, sec) & (deg, arcmin, arcsec) & (deg) &   \\
      \hline
      1E 0102-2-7219 & 1:05:55 & -72:04:21 & 1:05:53 & -72:04:51 & 16.5 & 4 \\
      \hline
    \end{tabular}
  }
  \setlength{\abovecaptionskip}{5pt} 
\caption{The table showcases data for a list of 8 SNR targets in the LMC and 1 SNR target in the SMC.  Column 1 includes the target name, column 2 and 3 specify the initial slit right ascension and declination respectively, column 4 and 5 specify the final slit right ascension and declination respectively, column 6 specifies the roll angle of the slit for these pointings, and column 7 identifies the number of pointings it takes to cover the extended source.}
\label{snr_list}
\end{table}

\begin{figure}[H]
\includegraphics[width=\textwidth]{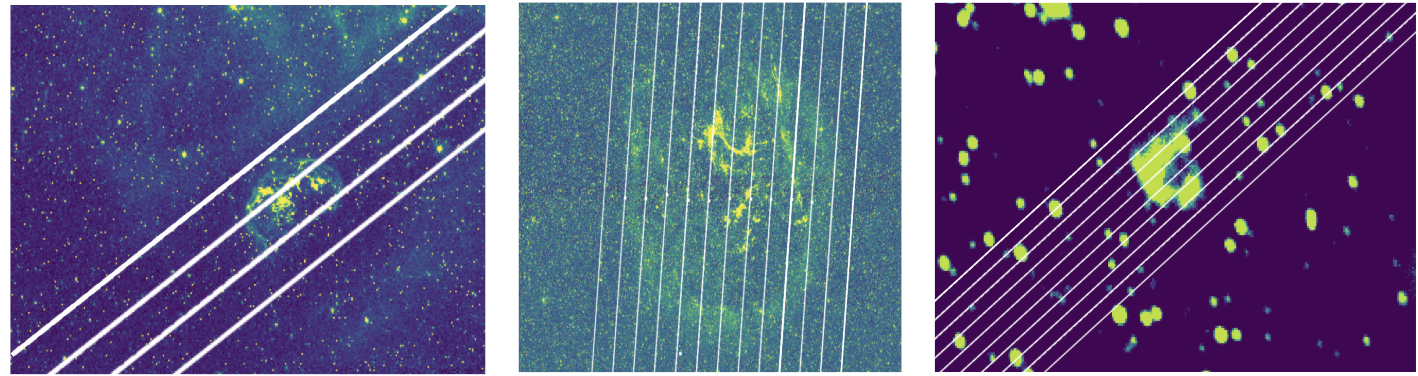}
\caption{The far left image depicts supernova remnant N103B with 3 SPRITE slits overlaid, the central image depicts supernova remnant N49 with 9 SPRITE slits overlaid, and the far right image depicts supernova remnant N132D with 12 SPRITE slits overlaid. Imaging data was extracted from HST.}
\label{pushbroom_maps}
\end{figure}

Figure \ref{pushbroom_maps} shows the push-broom mapping plan for three of the nine SNR targets mentioned in Table \ref{snr_list}.

\subsection{Data Validity}
SPRITE data products will produce spectral plots similar to those created using FUSE and IUE data.  With a bandwidth extending from 1000 to 1750\textup{~\AA}, SPRITE spectral observations will extend to higher energies than IUE and lower energies than FUSE.  We can therefore produce models that predict SPRITE spectral observations by combining data from FUSE and IUE.

The following analysis looks at a combination of FUSE and IUE data in supernova remnant LMC N132D to simulate what the SPRITE slit is likely to observe.  FUSE acquired 2 data products of this SNR, while IUE obtained 5 in the short-wavelength band.  Focusing on two IUE and two FUSE sources in N132D, two 1D spectral simulations were created to predict SPRITE observations.  Although the FUSE and IUE data products are not co-spatial, they are combined to give a rough estimate of what can be observed in the SPRITE bandpass.  The simulation validates SPRITE’s capability to resolve key emission lines.

\begin{figure}[H]
\includegraphics[width=\textwidth]{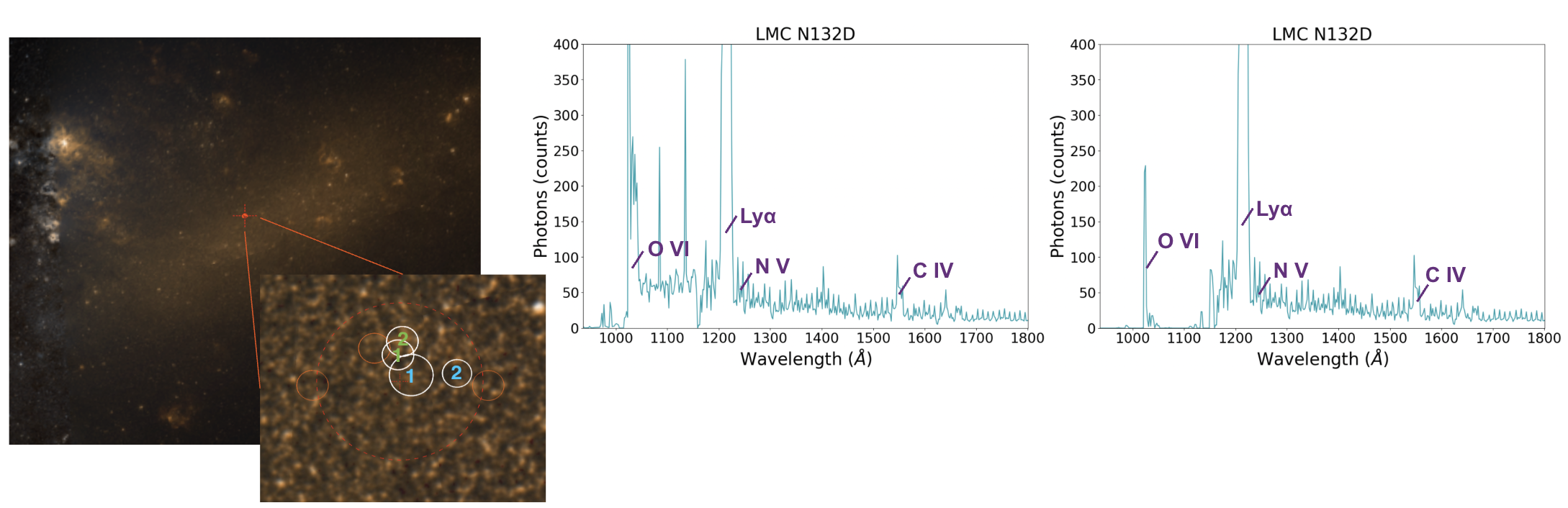}
\caption{The far left image taken from MAST shows four separate pointings within SNR N132D.  The green numbers label IUE pointings while blue numbers label FUSE pointings.  The central 1D spectral plot showcases N132D flux data using FUSE and IUE pointing 1 and 1.  The right 1D spectral plot showcases N132D flux data using FUSE and IUE pointing 2 and 2.}
\label{pointing_validity}
\end{figure}

Figure \ref{pointing_validity} models what SPRITE will observe using FUSE and IUE data, binned per \textup{\AA} to simulate SPRITE's spectral resolution.  Based on its high spectral resolution, especially compared to FUSE, SPRITE is expected to better resolve the emission lines visible in each of these models.  Both models show emission lines for O VI (1032, 1038\textup{~\AA}), N V (1238, 1242\textup{~\AA}), O IV] (1397, 1401\textup{~\AA}), Si IV (1394, 1403\textup{~\AA}), and C IV (1548, 1550\textup{~\AA}).  As mentioned previously, each emission line provides valuable spectral information that will be used to determine how gas and dust in the ISM is processed following the death of a massive star.

\section{Conclusions}

The SPRITE Mission aims to understand how the IGM and ISM became reionized by surveying high energy emission from diffuse gas to learn more about star formation.  SPRITE will collect spectral data between 1000 and 1750\textup{~\AA} with $\sim$ 10 arcsecond imaging resolution.

The mission will conduct three separate science surveys, SPIRES, SPRIGS, and STEMS, upon its launch in late 2025.  This paper produces a representative data product of what is expected of the SPRIGS and STEMS science surveys.  This data product is expected to be of the type generated by SPRIGS and STEMS science.  

SPRITE will employ push-broom mapping to survey emission from shocked gas around SNRs and star-forming regions in the Magellanic Clouds.  This technique will create 3D data cubes of recorded flux values for specified wavelength, right ascension, and declination.  SPRITE will also carry out a deep spectral mapping of local galaxies for an unprecedented picture of star formation and feedback processes that will be compared with existing surveys.

The SPRITE 12U CubeSat will image spectra on a 39 x 19 mm borosilicate glass MCP with $\sim$ 50 $\mu$m resolution.  The extended slit has a length of 1800 arcseconds and a width of 10 arcseconds. SPRITE boasts exceptional noise performance with a predicted on-orbit background of $\sim$ 0.5 counts \textup{cm$^{-2}$ s$^{-1}$}, not including stray light effects. 
 Additionally, the SPRITE effective area in the LyC is comparable to that of FUSE and HUT, two much larger and much more expensive instruments.

Models of 2D and 1D SNR spectra from the FUSE and IUE missions provide key insights into SPRITE spectral observations. The modeling process outlined in this paper discusses how spectral data from these two missions, in addition to imaging data from HST, was used to produce 2D and 1D spectral models spanning entire SNR in the Magellanic Clouds.  Using these models, we conclude that SPRITE data products will produce plots which showcase key emission lines such as O VI, N V, and C IV.  These emission lines are key shock tracers and will provide information on shock wind gas temperatures. 
 Spatially resolved ion maps produced by the SPRITE 3D data cubes will allow for derivation of gas cooling rates, dust destruction, and carbon depletion rates.  They will also provide information on the ion and electron temperatures in the shock and surrounding gas and dust, and how shock velocities move through gas and dust near the progenitor.

Commissioning and calibration targets were chosen on the basis of high S/N for given exposure times.  A list of 20 calibration star targets were produced by calculating the S/N per\textup{~\AA} and ensuring this value was above a threshold of 10.  Similarly, a list of 5 extended source commissioning targets were produced by calculating the S/N at key emission lines. 
 Extended source target pointings were determined using a given start and end right ascension and declination, and making an appropriate number of 10 arcsecond steps along a given roll angle.  The SPRITE mission is scheduled to launch in late 2025, with the first data products appearing on the MAST archive within approximately one year after commissioning is complete.

\section{Disclosures}

The authors declare that there are no financial interests, commercial affiliations, or other potential conflicts of interest that could have influenced the objectivity of this research or the writing of this paper.

\section{Code, Data, and Materials Availability}

The data supporting the findings of this article are proprietary and are not publicly available, but can be replicated via the published code. The archived version of the code described in this manuscript can be freely accessed through Github via \href{ https://github.com/SPRITE-CubeSat/SPRITE_DRP/tree/83835738a02e452b94b998181a09fdcc70c28920/simulated_observations}{SPRITE DRP}.

\section{Acknowledgments}
This work was funded by a grant from the National Aeronautics and Space Administration (NASA), Award no. 80NSSC19K0995 and 80NSSC24K0231, to the University of Colorado, Boulder.  The authors extend many thanks to all staff and students at LASP who have contributed their time and effort to the project. The authors wish to acknowledge the use of OpenAI’s ChatGPT (version August 2024) for assistance in language and grammar refinement. All AI-generated suggestions were reviewed, edited, and approved by the authors to ensure accuracy and integrity.

\appendix    


\bibliography{report}   
\bibliographystyle{spiejour}   

\listoffigures
\listoftables

\end{spacing}
\end{document}